\documentclass[prd,10pt,aps,showpacs,superscriptaddress,twocolumn,amsmath,amssymb,nofootinbib,longbibliography]{revtex4-1}

\usepackage{natbib}
\usepackage[colorlinks]{hyperref}
\hypersetup{linkcolor=blue,citecolor=blue}
\usepackage{orcidlink}
\usepackage[version=4]{mhchem}
\usepackage{graphicx}
\usepackage{comment}
\usepackage{dcolumn}
\usepackage{bm}
\usepackage{ulem}
\usepackage{color}
\usepackage{enumitem}
\usepackage{soul}

\usepackage{hyperref}

\begin{document}

\title{Analog event horizons from magnetoelectric materials}

\author{V. A. \surname{De Lorenci}\orcidlink{0000-0001-5880-2207}}
\email{delorenci@unifei.edu.br}
\affiliation{Instituto de F\'{\i}sica e Qu\'{\i}mica, Universidade Federal de Itajub\'a, \\
Itajub\'a, Minas Gerais 37500-903, Brazil}

\author{L. T. \surname{de Paula}\orcidlink{0009-0002-1192-3532}}
\email{tobias.l@ufabc.edu.br}
\affiliation{Centro de Matemática, Computação e Cognição, Universidade Federal do ABC, \\
Santo Andr\'e, S\~ao Paulo 09210-170, Brazil}
\affiliation{School of Mathematical Sciences, University of Nottingham,\\ University Park, Nottingham, NG7 2RD, UK}

\author{C. C. \surname{Holanda Ribeiro}\orcidlink{0000-0001-6571-4168}}
\email{caiocesarribeiro@alumni.usp.br}
\affiliation{International Center of Physics, Institute of Physics, University of Brasilia, 70297-400 Brasilia, Federal District, Brazil}
\affiliation{Instituto de F\'{\i}sica e Qu\'{\i}mica, Universidade Federal de Itajub\'a, \\
Itajub\'a, Minas Gerais 37500-903, Brazil}

\begin{abstract}
Analog models of gravity provide a laboratory setting to investigate curved space phenomena. In this context, linear magnetoelectric materials offer interesting possibilities for modeling such analog geometries. Here, general conditions under which a light ray enters a one-way (trapped) region, bounded by an analog event horizon, are identified. The results, thereby, establish linear magnetoelectric  materials as a new platform for analog black hole physics.

\end{abstract}

\maketitle

\setlength{\parskip}{0pt}
\section{Introduction}
\label{one}
Historically, the conceptual link between gravitational phenomena and the properties of continuous media can be traced back to Isaac Newton. Rather than proposing an analog equivalence, Newton speculated that fundamental gravitational interactions might be mediated by an underlying fluid-like aether \cite{burtt1924,10.1093/oxfordhb/9780199930418.013.23,erkul2025dispersionanaloguegravity}. Although this was an attempt to describe the physical mechanism of gravity itself rather than a formal analogy, it intuitively anticipated the mathematical connections between gravitational kinematics and fluid dynamics. The shift toward a true analog framework emerged in the early 20th century with Gordon's work on light propagation in material media \cite{gordon}. By recasting the refractive index as an effective geometric structure, his work provided one of the first systematic analog models, formally establishing a mathematical equivalence between material-media optics and spacetime geometry. Another seminal development occurred when Unruh demonstrated that linear perturbations in a moving fluid behave mathematically as massless scalar fields propagating in a curved spacetime, as encoded by the so-called acoustic metric \cite{Unruh1981}. Over the ensuing decades, these ideas have inspired a wide variety of studies, ranging from electromagnetic and optical media \cite{tamm,plebanski,plebanski2,novello,PhysRevA.89.053807,PhysRevA.104.043523,PhysRevA.93.053820} to acoustic, condensed matter, and fluid systems \cite{barcelo2,barcelo,MattVisser_1998,PhysRevA.76.033616,Jacobson2018,Garay,tedbhlaser}. Additionally, different effects have already been explored theoretically and experimentally in such systems, encompassing semiclassical phenomena such as Hawking radiation \cite{PhysRevLett.106.021302, steinhauer2016observation, Jeff2019, fabbri,PhysRevB.86.144505,PhysRevD.107.L121502}, as well as classical scattering problems such as superradiance \cite{PhysRevD.91.124018, Torres2017} and quasinormal mode decay \cite{PhysRevD.70.124006,PhysRevD.70.124032,PhysRevLett.125.011301,PhysRevD.111.104064}.

Currently, developments in analog gravity have increasingly focused on nonlinear and structured optical media, where strong light-matter interactions and engineered material properties enable the realization of effective spacetime geometries \cite{belgiorno2010hawking,ornigotti2018analog,smolyaninov2010gravity,hendi2020metamaterials}. In parallel, advances in the science and technology of optical materials, including metamaterials \cite{lapine2014} and magnetoelectric media \cite{fiebig,rivera,schmid,Tabares}, have opened new avenues for the investigation of analog models within an electromagnetic framework. In this context, magnetoelectric materials are particularly appealing, as they combine electric and magnetic responses in a single medium and provide additional degrees of freedom to tailor the effective metric experienced by light .

In this paper, we investigate analog models based on light propagation in magnetoelectric materials. The description is restricted to the regime of lossless and dispersionless systems, which consist of materials whose delay in their response to external electromagnetic perturbations is negligible.
Furthermore, we conduct the analysis within the regime of geometrical optics, where it is assumed that the total electromagnetic fields can be split in two contributions: a strong and slowly varying part, mainly responsible for activating the polarization and magnetization of the material, and a weak and rapidly varying field, which is the one that propagates in the medium.
An analog model based on materials whose magnetoelectric coefficient is antisymmetric is thus constructed, and the effective geometry is obtained. In particular, it is possible to show that one specific mode of propagation would experience a geometry presenting an analog event horizon.

In the next section, we briefly review the main aspects of light propagation in material media and obtain solutions for the phase velocities of waves in an antisymmetric magnetoelectric metamaterial. In particular, we show that one of the propagation modes experiences a trapped region for a specific choice of optical parameters.  Then, in Sec.~\ref{analog}, the effective geometry is explicitly derived, and it is shown that the resulting metric associated with an antisymmetric magnetoelectric system supports an analog event horizon. 
Final remarks are presented in Sec.~\ref{discussion}, and a comparison between the results obtained here and those using another possible representation for the constitutive relations \cite{depaula2025effectivespacetimedescriptionlight} is provided in Appendix~\ref{app}. 

Throughout the text, Greek indices run from 0 to 3 (spacetime indices) while Latin indices run from 1 to 3 (the three spatial directions) and the Einstein convention for summation is used, i.e.,  repeated indices in a monomial indicate summation. 
Partial and covariant derivatives with respect to coordinate $x^\mu$ are denoted, respectively, by a comma and a semi-comma followed by the corresponding $\mu$ index.

\section{Light ray propagation in linear magnetoelectrics}
\label{magnetoelectric}
Throughout this section, 3-dimensional component notation is being used, with the metric of the three-space in Galilean coordinates coinciding with the Kronecker delta $\delta_{ij}$. 
Therefore, without losing generality, we keep all indices at just one (lower) level, and Einstein's summation convention over repeated indices still applies.

When electric $E_i$ and magnetic $B_i$ fields are applied to an optical medium having magnetoelectric properties, polarization and magnetization phenomena may occur such that both fields can contribute to both effects. Let us introduce the auxiliary fields $D_i$ and $H_i$ by means of $D_i=\varepsilon_0 E_i +P_i$ and $H_i = (B_i/\mu_0) - M_i$. Restricting the analysis to the linear effects and neglecting spontaneous effects, the polarization and magnetization vectors will be given by \cite{rivera, w2y7-gfc8}
\begin{subequations}
\label{p&m}
\begin{align}
P_{i}&=\varepsilon_{0} \chi_{i j} E_{j}+\alpha_{i j} B_{j}, \\
M_{i}&=\frac{\tilde\chi_{i j} B_{j}}{\mu_{0}}+\alpha_{j i} E_{j},
\end{align}
\end{subequations}
where $\alpha_{ij}$ denotes the linear magnetoelectric coupling coefficient and is defined such that $\mu_0\alpha_{ij}$ has dimensions of ${\rm s}{\rm m}^{-1}$. Hereafter, it is assumed that the linear electric and magnetic susceptibility sectors are isotropic, i.e. $\chi_{ij} = \chi \delta_{ij}$ and $\tilde\chi_{ij} = \tilde\chi \delta_{ij}$, respectively. In this case,  the constitutive relations connecting the auxiliary fields to the fundamental electric and magnetic fields can be conveniently written as 
\begin{subequations}
\label{dh1}
\begin{align}
D_{i} & =\varepsilon E_{i} + \alpha_{i j} B_j,
\label{D}
\\
H_{i}&=\bar\mu B_{i}-\alpha_{j i} E_{j},
\label{H}
\end{align}
\end{subequations}
where the isotropic electric permittivity, $\varepsilon = \varepsilon_0(1+\chi)$, and the inverse magnetic permeability, $\bar\mu = (1-\tilde\chi)/\mu_0$, were defined. 

The analysis is henceforth specialized to the metamaterial regime characterized by $\varepsilon = -\epsilon$ and a purely antisymmetric magnetoelectric tensor, $\alpha_{ij} = -\alpha_{ji}$. Under the eikonal approximation, the probe fields are assumed to take the form $E_j \propto e_j \exp(i\Phi)$, where $e_j$ is the electric polarization vector and $\Phi$ is the rapidly varying phase defining the local wave vector $q_i \doteq \partial_i\Phi$ and angular frequency $\omega \doteq -\partial_t\Phi$. By specializing the general formulation previously reported in \cite{w2y7-gfc8}—which accounts for linear and nonlinear effects within the Boys-Post representation—to the present linear regime, the propagation problem reduces to the eigenvalue equation $Z_{ij}e_j = 0$, where the Fresnel tensor $Z_{ij}$ takes the form
\begin{equation}
    Z_{ij} = -\epsilon\delta_{ij}v^2 + \left( \epsilon_{ikl}\alpha_{jk} + \epsilon_{jkl}\alpha_{ik} \right) \kappa_l v - \bar{\mu} I_{ij},
    \label{zij}
\end{equation}
and where we defined the projector orthogonal to the wave vector, $I_{i j} \doteq \delta_{ij} - \kappa_{i} \kappa_{j}$, with $\kappa_i$ the $i$-th component of the unit wave-vector, $\kappa_i \doteq q_i / q$. It is not difficult to show that the magnitude of the phase velocities ($v = \omega/q$) will be obtained by solving ${\rm det }( Z_{ij} )=0$ for $v$, yielding
\begin{align}
v^{(1)}_\pm&=\frac{1}{\epsilon \mu}\left(\sigma_i\kappa_i\pm\sqrt{\sigma^2-\epsilon\mu}\right),
   \label{v1meta}\\
 v^{(2)}_\pm&=\frac{1}{\epsilon \mu}\left(\sigma_i\kappa_i\pm\sqrt{\left(\sigma_i\kappa_i\right)^2-\epsilon\mu}\right),
\label{v2meta}
\end{align}
where we have defined $\sigma_i \doteq \tfrac{\mu}{2}\epsilon_{i j k} \alpha_{j k}$, and $\sigma^2 = \sigma_i\sigma_i$.
%
This vector can be presented in terms of its Cartesian components as
\begin{align}
\vec \sigma = \mu(\alpha_{23},\alpha_{13},\alpha_{12}).
\label{vecsigma}
\end{align}

Now, let us assume that the permittivity is a function of the position, i.e., $\epsilon=\epsilon(\vec{x})$ such that the magnitude of the phase velocity changes along its path through the medium, and analyze each propagation mode separately. First, notice that the existence condition for the mode described in Eq.~(\ref{v1meta}) is $\sigma^2>\epsilon\mu$.  In this case, there will be three possible scenarios to consider:
\begin{enumerate}[label=--, leftmargin=*, widest=b]
    \item ~\textit{Non-recyprocal propagation.} This happens within the regions where $\mu\epsilon(\vec{x})<\sigma^2-(\sigma_i\kappa_i)^2$. In this case the material is nonrecyprocal, i.e., the modes will propagate in opposite directions with different phase velocities.
    
    \item ~\textit{Vanishing mode.} This scenario is defined by $\mu\epsilon(\vec{x})=\sigma^2-(\sigma_i\kappa_i)^2$. Only the positive solution will survive, the negative solution will freeze with null phase velocity.

  \item ~\textit{One-way region.} This corresponds to the regions where $\mu\epsilon(\vec{x})>\sigma^2-(\sigma_i\kappa_i)^2$. By inspecting Eq.~(\ref{v1meta}), one can see that, in this case, the square root term becomes smaller than the first one. Consequently, there will be two different solutions propagating in the same direction, leading to birefringence and a one-way propagation region for this particular mode.
\end{enumerate}

Notice that this behavior establishes distinct regions within the material, separated by the surfaces defined by 
\begin{equation}
\epsilon(\vec{x})=\frac{1}{\mu}[\sigma^2-(\sigma_i\kappa_i)^2].
\label{surface}
\end{equation}
In particular, the regions described by the third scenario already indicate the possibility of event-horizon-like structures forming within this class of materials.

Now, for the second mode given by Eq.~(\ref{v2meta}), since the square root term is always smaller than the first one, there will be just one possible scenario, which is a one-way propagation region. The existence condition for this solution is $\mu\epsilon<\left(\sigma_i\kappa_i\right)^2$.
It is important to notice that in order to guarantee the occurrence of all possible scenarios for both modes, one needs to combine the existence conditions of both solutions, leading to $\sigma^2<2(\sigma_i\kappa_i)^2$. Henceforth, we restrict our analysis to the parameter regime where only the existence condition for the first mode, given by Eq.~(\ref{v1meta}), is satisfied. By doing so, the second mode becomes evanescent, ensuring that only a single set of modes propagates through the medium.

\begin{figure}[h!]
\center
\includegraphics[width=\columnwidth]{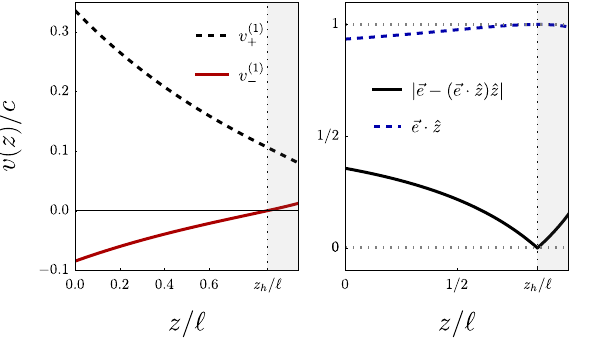}
\caption{Left: Phase velocities of light rays given by Eq.~(\ref{v1meta}), for wave propagation along the $z$-direction. The red solid line corresponds to $v^{\scriptscriptstyle(1)}_-$, while the black dashed corresponds to the $v^{\scriptscriptstyle(1)}_+$ solution . The material parameters are fixed to $\gamma=6\varepsilon_0$, $\mu=8\mu_0$, $\alpha_{12}=2\times10^{-3} {\rm A}{\rm V}^{-1}$, and $\alpha_{13}=\alpha_{23}=2.5\times10^{-3}{\rm A}{\rm V}^{-1}$. The dotted vertical line indicates the position of the surface determined by Eq.~(\ref{horizonposition}), located at $z/\ell \simeq 0.86$. At this boundary, the $v^{\scriptscriptstyle(1)}_-$ mode vanishes and subsequently changes sign, signaling the transition to the other region (in gray). Beyond this point, propagation is restricted to modes traveling to the right and the material becomes birefringent, highlighting the one-way character of wave propagation. This behavior is a signature of the horizon-like nature of the system, as experienced by this mode. Right: Behavior of the polarization vector linked to mode $v^{\scriptscriptstyle(1)}_-$. Notice that the vector aligns with the direction of propagation at the boundary.}
\label{figvel}
\end{figure}

Let us now explore a specific case where we set the electric permittivity as follows
\begin{equation}
    \epsilon(z)= \gamma {\rm e}^{z/\ell},
    \label{model}
\end{equation}
where $\ell$ is the width of the slab and $\gamma$ is a parameter with dimensions of electric permittivity that can be conveniently adjusted. Moreover, we also set the propagation in the direction of $z$. In this case, the surface dividing the material is located at $z=z_h$, given by
\begin{equation}
    \frac{z_h}{\ell} = \ln \left[ \frac{\mu}{\gamma} (\alpha_{23}^2 + \alpha_{31}^2) \right].
    \label{horizonposition}
\end{equation}
Notice that the existence of such boundary is solely due to the magnetoelectric effect. The behavior of the phase velocities given by Eq.~(\ref{v1meta}) is illustrated in Fig.~\ref{figvel} for a certain set of optical coefficients. Initially, in the white region, the medium is nonreciprocal, supporting counter-propagating modes with different speeds. As the wave propagates along the $z$-axis, the increase of the permittivity effectively acts as an increasing optical density, which suppresses the intrinsic propagation speed of both modes. Physically, this kinematic evolution behaves analogously to a wave propagating in a moving fluid where the local speed of sound progressively drops. Eventually, this effective drag completely overpowers the diminishing backward speed, causing the $v^{\scriptscriptstyle(1)}_-$ mode to vanish at the boundary given by Eq.~(\ref{horizonposition}). This critical balance resembles the mode-freezing phenomenon characteristic of an event horizon in general relativity. Past this boundary, in the gray region, $v^{\scriptscriptstyle(1)}_-$ changes sign, meaning all perturbations are strictly swept forward. Consequently, the medium becomes birefringent in this one-way regime, as both solutions propagate in the same direction with distinct velocities. This mode-trapping mechanism anticipates the emergence of an analog event horizon, whose geometrical aspects are discussed in Sec.~\ref{analog}.

For completeness, we also show in Fig.~\ref{figvel}, right panel, the curious behavior of the polarization vector, $\vec{e}$, associated with the negative mode in Eq.~(\ref{v1meta}). As the boundary is approached, the mode $v^{\scriptscriptstyle(1)}_-$ vanishes. In this limit, the polarization vector becomes aligned with the wave vector, suggesting that interesting physics involving energy and momentum fluxes occur near the boundary due to magnetoelectric effects. These aspects will be explored in a future work. 

\section{Analog models}
\label{analog}
In what follows, the effective metric perceived by the light rays propagating in an antisymmetric magnetoelectric medium is obtained. For notational simplicity, we adopt units where $c=1$. Furthermore, we assume the background spacetime as described by the Minkowski metric, which, in Cartesian coordinates, reads $\eta_{\mu\nu} = {\rm diag}(1, -1, -1, -1)$. 
Referring back to the definitions above Eq.~\eqref{zij}, we define the wave 4-vector $k_{\mu} \doteq (\omega, -\vec q\,)$, such that $k^2 = \omega^2-q^2$, with $q = \|\vec q\,\|$. We introduce the 4-vector $V^\mu$ representing the velocity field of a family of observers at rest relative to the optical material through which light is propagating. In the case of our interest, the observers will always be at rest in the laboratory frame, i.e., $V^\mu = \delta^\mu_0$. In terms of such velocity field, $\omega = k_\mu V^\mu$. Furthermore, as $k^2 = k_\mu k_\nu \eta^{\mu\nu}$, it follows that 
$q^2 = -k_\mu k_\nu h^{\mu\nu}$, where the projector in the three-dimensional space section is defined as
\begin{equation}
    h^{\mu\nu} = \eta^{\mu\nu}-V^\mu V^\nu,
\nonumber
\end{equation}
such that $h_{\mu\nu}V^\nu=0$ and $h_{\mu\alpha}h^{\alpha}{}_{\nu} = h_{\mu\nu}$. We thus conveniently define the four-vector 
\begin{align}
q^\mu = h^{\mu\nu} k_\nu = k^\mu-\omega V^\mu = (0,\vec q).
\nonumber
\end{align}
Notice that $q_\mu q^\mu=-q^2$. Finally, we define the four-vector $\sigma^\mu= (0, \vec{\sigma})$, where $\vec\sigma$ is defined by Eq.~(\ref{vecsigma}).

Let us examine a possible model based on the solution in Eq.~(\ref{v1meta}).
Squaring the phase velocity solution, one obtains that
\begin{align}
    \omega^2-2v_o^2\omega\sigma_iq_i+v_o^4(\sigma_i q_i)^2-v_M^2q^2=0,
    \label{dispersionrel}
\end{align}
where $v_o^2\doteq(\epsilon\mu)^{-1}$ and $v_M^2\doteq v_o^4(\sigma^2-\epsilon\mu)$. Note that the dispersion relation above makes no reference to the coordinates used in the spatial sector of the (flat) spacetime. Thus, in order to identify an interesting family of analog spacetimes, let us orient the $z$ axis to coincide with the direction of $\vec{q}$,
i.e., the wave-vector reduces to $\vec{q}=(0, 0, q_3)$. Now, using that $\sigma_iq_i=\sigma_3q_3=-\sigma_3{}^\mu k_\mu$, where $\sigma_3{}^\mu=\sigma_3\delta^\mu_3$, and $q^2=-k^2+\omega^2$, one can recast Eq.~(\ref{dispersionrel}) in the form  $ g^{\mu\nu} k_\mu k_\nu = 0$, with
\begin{equation}
    g^{\mu\nu}=\eta^{\mu\nu}-\left(1-v_M^2\right)h^{\mu\nu}+2v_o^2V^{(\mu}\sigma_3{}^{\nu)}+v_o^4\sigma_3{}^{\mu}\sigma_3{}^{\nu},
\label{metriccontra}
\end{equation}
where the parentheses encompassing two indices denote symmetrization with unit weight. The tensor $g_{\mu\nu}$, defined by $g_{\mu\nu}g^{\nu\alpha}=\delta^{\alpha}_\mu$, reads
\begin{equation}
\begin{split}
    g_{\mu\nu}=\frac{1}{v_M^2}&\left\{\eta_{\mu\nu} -2v_o^2 \sigma_{3(\mu} V_{\nu)} \right.\\
    &\left.-\left[1-v_o^4(\sigma_1^2+\sigma_2^2-\epsilon\mu)\right]V_\mu V_\nu\right\},
    \end{split}
    \label{effectivemetric}
\end{equation}
and it represents an effective metric tensor built from the material optical properties.

The spacetime described by $g_{\mu\nu}$ is such that light rays propagating in the $z$ direction are indistinguishable from light rays propagating in the analog model. In general, light rays in curved spaces are determined by $ds^2=g_{\mu\nu}dx^\mu dx^\nu=0$, and, for $dx=dy=0$,  one finds
\begin{equation}
v_o^4(\sigma_1^2+\sigma_2^2-\epsilon\mu)dt^2+2v_o^2\sigma_3dzdt-dz^2=0.
    \label{analogmodel}
\end{equation}

Analogously to Sec.~\ref{magnetoelectric}, we assume $\epsilon=\epsilon (z)$, and determine the general conditions for which the analog spacetime contains event horizons.
Given that the metric coefficients are independent of the coordinate $t$, the spacetime possesses a stationary killing vector field defined by $\chi^\mu = (1,0,0,0)$.  Following the standard treatment of stationary spacetimes \cite{carroll2004spacetime}, a killing horizon is defined as a hypersurface where the norm of the killing vector vanishes, $g_{\mu\nu}\chi^\mu \chi^\nu = 0$, provided the surface is null. Calculating the norm directly from the metric gives
\begin{equation}
     g_{\mu\nu}\chi^\mu\chi^\nu = g_{00} = \frac{\sigma_1^2 + \sigma_2^2 - \epsilon \mu}{\sigma^2-\epsilon\mu}.
\end{equation}
Then, the condition $g_{00} = 0$ implies that horizons, if present, are  located at coordinates $z=z_h$ satisfying $\epsilon(z_h) = (\sigma_1^2 + \sigma_2^2)/\mu$, which is equivalent to Eq.~(\ref{surface}) when the propagation is set in $z$ direction.

To show that the hypersurface $z = z_h$ is indeed null, we examine the normal one-form $w_\mu = \partial_\mu(z - z_h) = (0, 0, 0, 1)$. The condition for a null surface requires $g^{\mu\nu}w_\mu w_\nu = g^{33} = 0$. Using the inverse metric components (\ref{metriccontra}) one gets
\begin{equation}
    g^{33} = \frac{g_{00}}{g_{00}g_{33} - g_{03}^2}.
\end{equation}
It follows that $g^{33} = 0$ precisely when $g_{00} = 0$, provided the determinant $ g_{00}g_{33} - g_{03}^2$ remains non-vanishing. Consequently, this condition defines the location of event horizons in this analog model.
As we examined, past this point no light will be able to propagate backwards, perturbations will be dragged by optical effects associated with the magnetoelectric properties of the medium. 

\begin{figure}[h!]
\center
\includegraphics[width=\columnwidth]{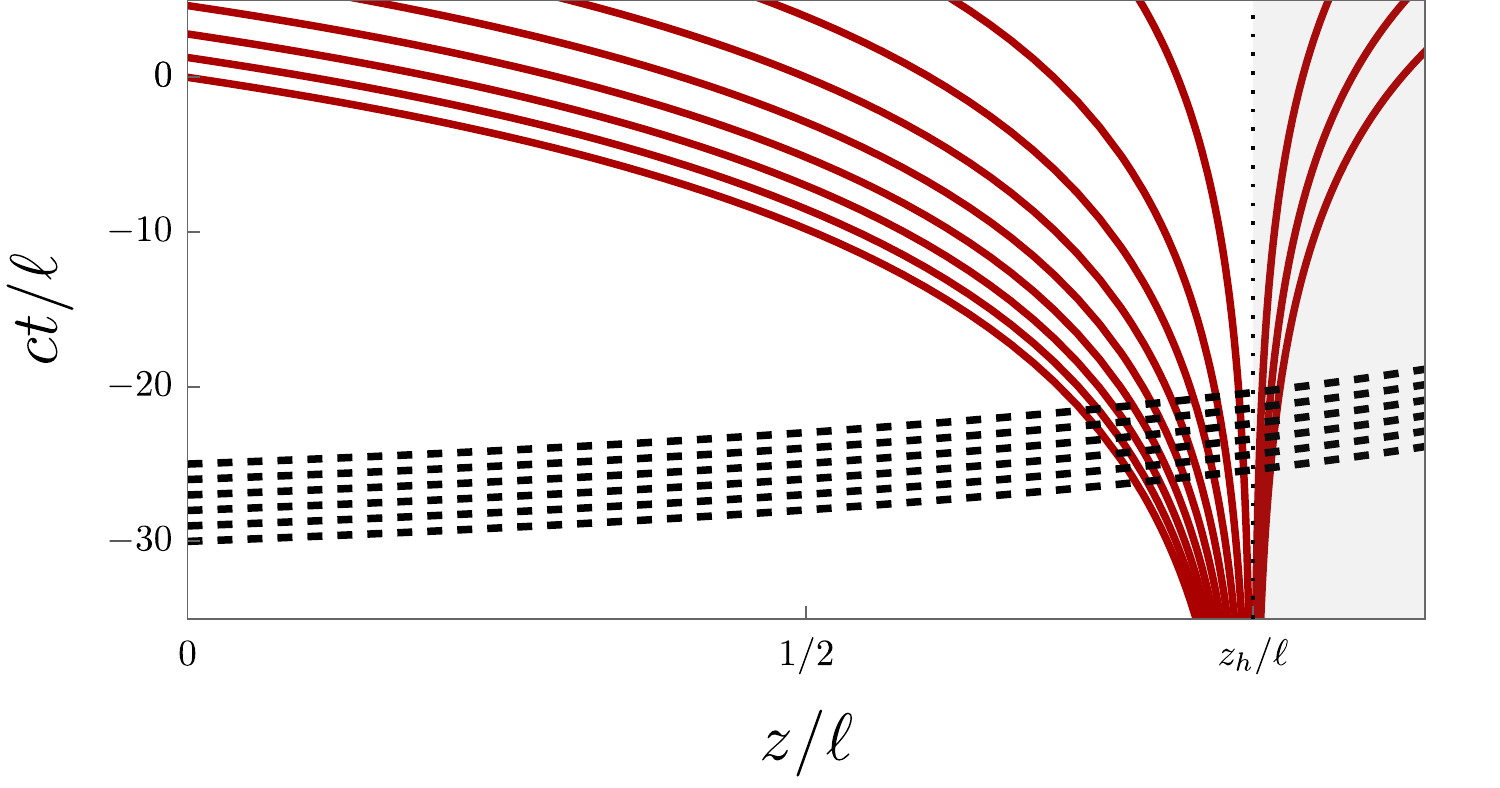}
\caption{Worldline of light rays in the analog model described by Eq.~(\ref{analogmodel}). Continuous lines correspond to rays whose phase velocity vanish at the horizon ($z=z_h$) and propagate away from it, whereas the dashed curves depict rays that fall directly into the analog black hole. Parameters are the same as in Fig.~\ref{figvel}. One should note the similarity between these geodesics and those associated with the Schwarzschild black hole spacetime.}
\label{fig2}
\end{figure}

To illustrate this phenomenon, consider again the particular model described by Eq.~(\ref{model}). Using Eq.~(\ref{analogmodel}), we determine the worldlines of light rays by solving for dt/dz, as shown in the left panel of Fig.~\ref{fig2} for the same material parameters as in Fig.~\ref{figvel}. Two possible solutions are found, corresponding to whether the phase velocity has a plus or minus sign in Eq.~\eqref{v1meta}, and one readily sees that the surface $z=z_h$ is indeed an analog event horizon, as it can only be crossed by light rays in one direction, i.e. light rays at $z>z_h$ cannot access the region $z<z_h$. 

\section{Final remarks}
\label{discussion}
The emergence of horizon-like behavior in linear magnetoelectrics reveals that nonreciprocity, when combined with other optical coefficients, and properly engineered, can mimic the causal structure associated with curved spacetimes. This observation places these materials among the simplest electromagnetic systems capable of reproducing stationary analog metrics, without requiring nonlinearity or background flow. The antisymmetric magnetoelectric coupling acts as an effective drag that affects light, producing mixed time-space components in the metric and establishing conditions under which light rays become trapped or unidirectional.

A requirement for the emergence of the trapped mode is a negative electric permittivity ($\varepsilon < 0$), which has been tailored, for example, in \ce{BaTiO3\text{/}Y3Fe5O12} (YIG) composites \cite{Wang_2017}. The realization of an antisymmetric magnetoelectric response is an additional ingredient that is relevant from an experimental perspective. Nevertheless, within the constitutive framework adopted here, the magnetoelectric sector emerges naturally as a contribution that is decoupled from the purely dielectric and magnetic responses, facilitating a material design that satisfies the conditions for establishing a horizon.

 Future developments may address the inclusion of dispersion, dissipation, and nonlinear corrections to evaluate their impact on the effective geometry and to investigate possible analogs of quantum field effects, such as Hawking-like emission. In particular, exploring the spatial dependencies of the magnetoelectric coefficients could also allow for the emulation of rotation-like effects, establishing formal analogs with rotating black hole geometries. Hence, the linear magnetoelectric model proposed here not only clarifies the geometric role of magnetoelectricity, but also opens a new direction for exploring the kinematic aspects of general relativity within the science of optical materials.

\begin{acknowledgments}
LTP thanks A. Fabbri for his kind hospitality at the University of Valencia, where  preliminary stages of this work were developed, and for valuable discussions on analog event horizons. This work was partially supported by the Conselho Nacional de Desenvolvimento Cient\'{\i}fico e Tecnol\'ogico (CNPq, Brasil), Grant 302492/2022-4, by the São Paulo Research Foundation (FAPESP), Brasil - Processes Number 2023/07013-2 and 2025/17924-8, and by the Funda\c{c}\~ao de Apoio \`a Pesquisa do Distrito
Federal, Grant 00193-00002051/2023-14.
\end{acknowledgments}

\appendix

\section{Tellegen vs Boys-Post representation}
\label{app}
It is instructive to compare the results obtained in the present manuscript with those reported in previous work~\cite{depaula2025effectivespacetimedescriptionlight}. Both analyses investigate electromagnetic wave propagation in magnetoelectric media with antisymmetric magnetoelectric response and isotropic linear electric and magnetic sectors. The physical effects identified in each case are closely related; however, they are not expected to coincide. This is because, when examined in detail, the underlying material systems are not the same.

The origin of this distinction lies in the representation adopted to describe the electromagnetic response of the medium. In Ref.~\cite{depaula2025effectivespacetimedescriptionlight}, the material was characterized directly through the constitutive relations between the fields, which effectively correspond to choosing the fields $\mathbf{E}$ and $\mathbf{H}$ as thermodynamic variables in the free-energy density, whereas in the present analysis these variables are chosen to be the fundamental fields $\mathbf{E}$ and $\mathbf{B}$. These choices correspond, respectively, to the Tellegen and Boys-Post representations~\cite{2010eabf.book.....M,wernerwein}. It is well known that these representations, although formally related, lead to different constitutive parameterizations and therefore to different material classes when specific symmetry assumptions are imposed.

The constitutive parameters defined within the two representations are related by
\begin{align}
\varepsilon_{ij}^{\scriptscriptstyle \boldsymbol{E\!B}} &=
\varepsilon_{ij}^{\scriptscriptstyle \boldsymbol{E\!H}}
- \alpha^{\scriptscriptstyle \boldsymbol{E\!H}}_{il}
\,\bar\mu^{\scriptscriptstyle \boldsymbol{E\!H}}_{lk}\,
\alpha^{\scriptscriptstyle \boldsymbol{E\!H}}_{jk},
\nonumber\\
\alpha^{\scriptscriptstyle \boldsymbol{E\!B}}_{ij} &=
\bar\mu^{\scriptscriptstyle \boldsymbol{E\!H}}_{jk}
\,\alpha^{\scriptscriptstyle \boldsymbol{E\!H}}_{ik},
\nonumber\\
\mu^{\scriptscriptstyle \boldsymbol{E\!B}}_{ij} &=
\mu^{\scriptscriptstyle \boldsymbol{E\!H}}_{ij},
\end{align}
which explicitly show that the electric, magnetic, and magnetoelectric sectors are intertwined when moving between the two descriptions.

In the present work, the linear electric and magnetic responses are chosen to be isotropic,
\begin{equation}
\mu^{\scriptscriptstyle \boldsymbol{E\!B}}_{ij} = \mu\,\delta_{ij},
\qquad
\varepsilon^{\scriptscriptstyle \boldsymbol{E\!B}}_{ij} = \varepsilon\,\delta_{ij},
\end{equation}
while the linear magnetoelectric tensor is assumed to be antisymmetric,
\begin{equation}
\alpha^{\scriptscriptstyle \boldsymbol{E\!B}}_{ij}
= \epsilon_{ijk}\,\alpha_k,
\qquad
\alpha_i = (\alpha_1,\alpha_2,\alpha_3).
\end{equation}
Under these assumptions, the corresponding Tellegen electric permittivity becomes
\begin{equation}
\varepsilon^{\scriptscriptstyle \boldsymbol{E\!H}}_{ij}
= \delta_{ij}\bigl(\varepsilon + \mu\,\alpha^2\bigr)
+ \mu\,\alpha_i\alpha_j,
\label{eq:epsEH_aniso}
\end{equation}
which is intrinsically anisotropic. Therefore, although the present model is isotropic in its linear electric and magnetic sectors when formulated in the Boys-Post representation, it corresponds, from the Tellegen perspective, to a material with an intrinsically anisotropic electric response. As a result, the medium studied here is not equivalent to the one considered in ~\cite{depaula2025effectivespacetimedescriptionlight}, where isotropy was imposed directly on the Tellegen constitutive parameters. It is thus natural that the two analyses lead to different effective descriptions of wave propagation. 

This observation highlights an important conceptual advantage of the Boys-Post formulation. In this representation, the magnetoelectric coupling responsible for the horizon-like behavior can be introduced without being intertwined with the linear dielectric and magnetic sectors. As a result, the physical mechanism leading to the emergence of an effective spacetime geometry is clearly isolated and can be traced directly to the antisymmetric magnetoelectric response. In contrast, within the Tellegen description, the same mechanism becomes obscured by the unavoidable mixing between magnetoelectric and dielectric parameters.

From a practical standpoint, this distinction is also relevant. While the Boys-Post description points to a relatively simple material architecture capable of supporting horizon-like phenomena, the equivalent Tellegen medium would require a finely tuned anisotropic electric response, which is substantially more demanding from an experimental and material-engineering perspective. In this sense, the Boys-Post representation does not merely offer an alternative parametrization, but rather provides a more natural and efficient framework for identifying and designing magnetoelectric systems that realize analog gravity scenarios. These considerations are consistent with previous studies of electromagnetic wave propagation in magnetoelectric media, including investigations of nonlinear optical effects formulated within the Boys-Post representation \cite{w2y7-gfc8}, where the analysis has been shown to be more direct and physically transparent than in the corresponding Tellegen description. 

\bibliography{ref}
\end{document}